\documentclass[%
preprint,
amsmath,amssymb,
aps,
]{revtex4-1}

\usepackage{graphicx} 
\graphicspath{ {Figures/} }
\usepackage{epstopdf}
\usepackage[caption=false]{subfig}
\usepackage{amsmath}
\usepackage{amsfonts}
\usepackage{amssymb}
\usepackage{bm}
\usepackage{hyperref}
\usepackage{amsthm}
\usepackage{color}

\usepackage{mathrsfs}
\usepackage[vlined,ruled]{algorithm2e}
\usepackage{url}
\usepackage{color}
\usepackage{algorithmic}
\usepackage{bbm}
\usepackage{booktabs}
\usepackage{array}
\usepackage[table]{xcolor}
\usepackage{yfonts}

\DeclareMathOperator*{\argmin}{argmin}

\newcommand\oprocendsymbol{\hbox{$\square$}}
\newcommand\oprocend{\relax\ifmmode\else\unskip\hfill\fi\oprocendsymbol}

\begin{document}
\bibliographystyle{naturemag}
\title{Teaching Recurrent Neural Networks to Modify Chaotic Memories by Example}
\author{Jason Z. Kim}
\affiliation{Department of Bioengineering, University of Pennsylvania, Philadelphia, PA, 19104}
\author{Zhixin Lu}
\affiliation{Department of Bioengineering, University of Pennsylvania, Philadelphia, PA, 19104 }
\author{Erfan Nozari}
\affiliation{Department of Electrical and Systems Engineering, University of Pennsylvania, Philadelphia, PA, 19104 }
\author{George J. Pappas}
\affiliation{Department of Electrical and Systems Engineering, University of Pennsylvania, Philadelphia, PA, 19104}
\author{Danielle S. Bassett}
\affiliation{Departments of Bioengineering, Physics \& Astronomy, Electrical \& Systems Engineering, Neurology, and Psychiatry, University of Pennsylvania, Philadelphia, PA, 19104}
\affiliation{Santa Fe Institute, Santa Fe, NM 87501}
\affiliation{To whom correspondence should be addressed: dsb@seas.upenn.edu}
\date{\today}

\begin{abstract}
	The ability to store and manipulate information is a hallmark of computational systems. Whereas computers are carefully engineered to represent and perform mathematical operations on structured data, neurobiological systems perform analogous functions despite flexible organization and unstructured sensory input. Recent efforts have made progress in modeling the representation and recall of information in neural systems. However, precisely how neural systems learn to modify these representations remains far from understood. Here we demonstrate that a recurrent neural network (RNN) can learn to modify its representation of complex information using only examples, and we explain the associated learning mechanism with new theory. Specifically, we drive an RNN with examples of translated, linearly transformed, or pre-bifurcated time series from a chaotic Lorenz system, alongside an additional control signal that changes value for each example. By training the network to replicate the Lorenz inputs, it learns to autonomously evolve about a Lorenz-shaped manifold. Additionally, it learns to continuously interpolate and extrapolate the translation, transformation, and bifurcation of this representation far beyond the training data by changing the control signal. Finally, we provide a mechanism for how these computations are learned, and demonstrate that a single network can simultaneously learn multiple computations. Together, our results provide a simple but powerful mechanism by which an RNN can learn to manipulate internal representations of complex information, allowing for the principled study and precise design of RNNs.
\end{abstract}
\maketitle

\section{Introduction}
Computers analyze massive quantities of data with speed and precision \cite{nvidia2020,intel2019}. At both the hardware and software levels, this performance depends on fixed and precisely engineered protocols for representing and executing basic operations on binary data \cite{intel2019,Neumann1945,Alglave2008}. In contrast, neurobiological systems are characterized by flexibility and adaptability. At the biophysical level, neurons undergo dynamic changes in their composition and patterns of connectivity \cite{Zhang2011,Faulkner2008,Dunn2012,Craik2006}. At the cognitive level, they abstract spatiotemporally complex sensory information to recognize objects, localize spatial position, and even control new virtual limbs through experience \cite{Tacchetti2018,Moser2008,Ifft2013}. Hence, neural systems appear to work on fundamentally different computing principles that are learned, rather than engineered.

To uncover these principles, artificial neural networks have been used to study the representation and manipulation of information. While feed-forward networks can classify input data \cite{Sainath2015}, biological organisms contain recurrent connections that are necessary to sustain short-term memory of internal representations \cite{Jarrell2012}, allowing for more complex functions such as tracking time, distance, and emotional context \cite{Lee2015,Wang2018,Weber2017,Burak2009,Yoon2013}. Further, recurrent neural systems actually manipulate internal representations to simulate the outcome of dynamic processes such as kinematic motion and navigation \cite{Hegarty2004,Kubricht2017,Pfeiffer2013}, and to decide between different actions \cite{Gold2007}. How do recurrent neural systems learn to represent and manipulate complex information?

One promising line of work involves representing static memories as patterns of neural activity, or \emph{attractors}, to which a network evolves over time \cite{Strogatz1994}. These attractors can exist in isolation (e.g. an image of a face) or as a continuum (e.g. smooth translations of a face) using Hopfield or continuous attractor neural networks (CANNs), respectively \cite{Yang2017,Wu2016}. Other studies use a differentiable neural computer (DNC) to read and write information to these attractor neural networks to solve complex puzzles \cite{Graves2016}. For understanding neurobiological systems, these memory networks are limited by requiring specifically engineered patterns of connectivity, and cannot manipulate time-varying memories necessary to plan and produce speech and music \cite{Carroll2004,Fee2010,Donnay2014}. Additionally, DNCs artificially segregate the computing and storage components. Hence, we seek a single neural system that learns to both represent and manipulate temporally complex information by perceiving and replicating examples.

In this work, we use the \emph{reservoir computing framework} \cite{Qiao2017} to obtain such a system (the reservoir), where the complex information is a chaotic attractor that is not static, but evolves in a deterministic yet unpredictable manner through time \cite{Lorenz1963}. Prior work has demonstrated the reservoir's ability to represent and switch between isolated attractors by imitating examples \cite{Jaeger2010,Sussillo2009}. Here, we demonstrate that reservoirs can further learn to interpolate and extrapolate translations, linear transformations, and even bifurcations on their representations of chaotic attractor manifolds simply by imitating examples. Further, we put forth a mechanism of how these computations are learned, providing insights into the set of possible computations, and offering principles by which to design effective networks.

\section{Mathematical Framework}
\begin{figure}[h!]
	\includegraphics[width=90mm]{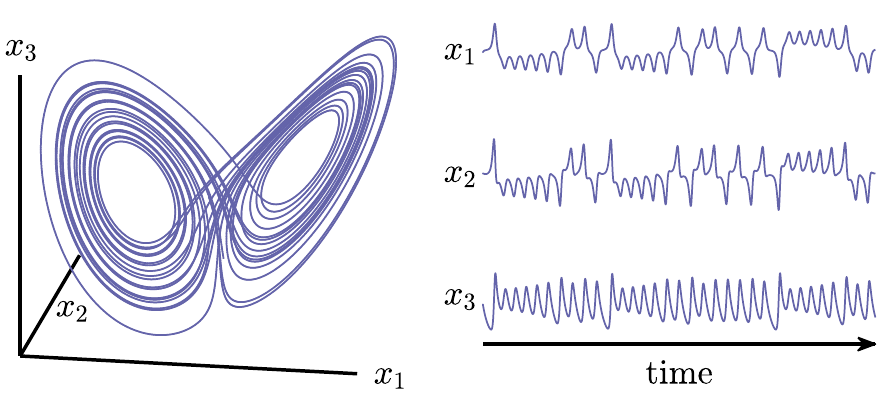}
	\caption{\textbf{Chaotic Lorenz manifold.} Lorenz attractor plotted in space (left) and time (right).}
	\label{fig:f1}
\end{figure}

Neural systems represent and manipulate periodic stimuli through example, such as baby songbirds modifying their song to imitate adult songbirds \cite{Fee2010}. However, they also perform more advanced and original manipulations on aperiodic stimuli with higher-order structure, such as musicians improvising on jazz melodies \cite{Donnay2014}. To model such complex stimuli, we use chaotic attractors that evolve deterministically yet unpredictably along a global structure: a fractional-dimensional manifold. Specifically, we consider the Lorenz attractor defined as
\begin{equation}
\label{eq:lorenz}
\begin{aligned}
\dot{x}_1 &= \sigma(x_2-x_1)\\
\dot{x}_2 &= x_1(\rho - x_3) - x_2\\
\dot{x}_3 &= x_1x_2 - \beta x_3,
\end{aligned}
\end{equation}
and use the parameters $\sigma = 10, \beta = 8/3, \rho = 28$ from the original study \cite{Lorenz1963} (Fig.~\ref{fig:f1}).

Next, we model the neural system as a recurrent neural network driven by our inputs
\begin{align*}
\frac{1}{\gamma} \dot{\bm{r}} = -\bm{r} + \bm{g}\left(A\bm{r} + B\bm{x} + \bm{d}\right),
\end{align*}
where $\bm{r}$ is a real-valued vector of $N$ reservoir neuron states, $A$ is an $N \times N$ matrix of inter-neuron connections, $B$ is an $N \times 3$ matrix of connections from the inputs to the neurons, $\bm{d}$ is an $N \times 1$ bias vector, $\bm{g}$ is a scalar activation function applied entry-wise to its input arguments (hence mapping vectors to vectors), and $\gamma$ is a time constant. 

\begin{figure}[h!]
	\centering
	\includegraphics[width=6.5in]{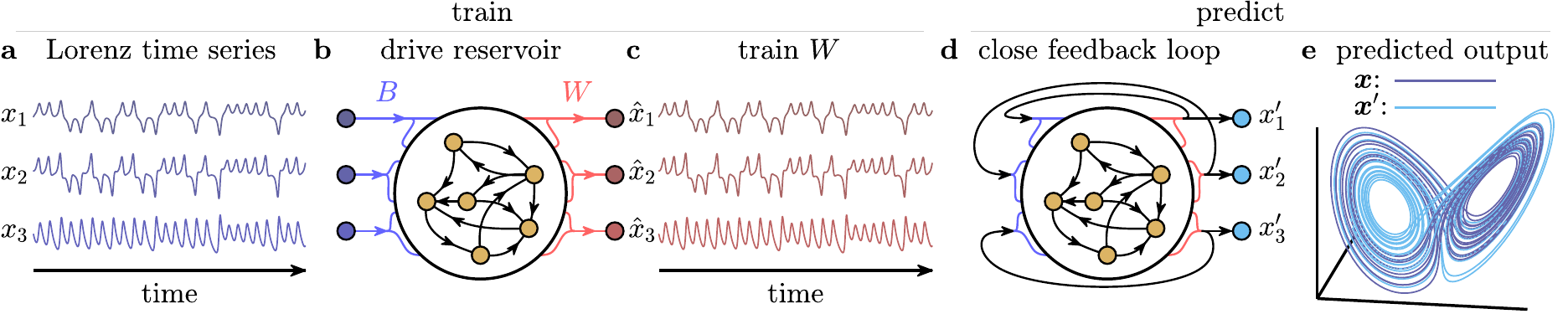}
	\caption{\textbf{Representing chaotic attractors with reservoirs.} (\textbf{a}) Time series of a chaotic Lorenz attractor that (\textbf{b}) drives the recurrent neural network reservoir. (\textbf{c}) Weighted sums of the reservoir states are trained to reproduce the original time series. (\textbf{d}) By using these weighted sums of reservoir states to drive the reservoir instead of the inputs, (\textbf{e}) the reservoir autonomously evolves along a trajectory that projects to a Lorenz-shaped chaotic manifold.}
	\label{fig:f2}
\end{figure}

Several prior studies use echo state \cite{Jaeger2010} and FORCE learning \cite{Sussillo2009} which allow reservoirs to predict a chaotic time series by modifying the inter-neuron connections. This modification can be accomplished by using the chaotic time series $\bm{x}(t)$ to drive the reservoir, thereby generating the reservoir time series $\bm{r}(t)$ (Fig.~\ref{fig:f2}a,b). Here, $\bm{x}(t)$ and $\bm{r}(t)$ are $3 \times T$ and $N \times T$ matrices, respectively, from numerically evolving the differential equations over $T$ time steps. By solving for a simple $3 \times N$ readout matrix $W$ that uses linear combinations of reservoir states to approximate the input by minimizing the matrix 2-norm (see Supplement)
\begin{align*}
W = \argmin_W\|W\bm{r}(t) - \bm{x}(t)\|_2,
\end{align*}
the output $\hat{\bm{x}}(t) = W\bm{r}(t)$ mimics the input $\bm{x}(t)$ (Fig.~\ref{fig:f2}c). Finally, we close the feedback loop by substituting the output as the input to create the autonomous reservoir (Fig.~\ref{fig:f2}d)
\begin{align*}
\frac{1}{\gamma} \dot{\bm{r}}' = -\bm{r}' + \bm{g}\left((A+BW)\bm{r}' + \bm{d}\right),
\end{align*}
whose evolution projects to a Lorenz-shaped attractor as $\bm{x}'(t) = W\bm{r}'(t)$ (Fig.~\ref{fig:f2}e). Hence, reservoirs sustain representations of complex temporal information by learning to autonomously evolve along a chaotic attractor from example inputs.

To study how reservoirs might perform computations by modifying the position or geometry of these representations in a desired way, we first adapt the framework to include a vector of control parameters $\bm{c}$ that map to the reservoir neurons through matrix $C$ to yield
\begin{align*}
\frac{1}{\gamma} \dot{\bm{r}} = -\bm{r} + \bm{g}\left(A\bm{r} + B\bm{x} + C\bm{c} + \bm{d}\right).
\end{align*}
Such control parameters were also previously used to switch between multiple attractor outputs \cite{Sussillo2009}. The second adaptation is to approximate the reservoir dynamics using a Taylor series to quadratic order around equilibrium values $\bm{r}^*,\bm{x}^*=\bm{0},\bm{c}^*=\bm{0}$, yielding
\begin{align}
\label{eq:reservoirc}
\frac{1}{\gamma}\delta\dot{\bm{r}} = -\delta\bm{r} + U(A\delta\bm{r} + B\bm{x} + C\bm{c}) + V(A\delta\bm{r} + B\bm{x} + C\bm{c})^2.
\end{align}
Here, $\delta\bm{r} = \bm{r}-\bm{r}^*$, $U$, and $V$ are diagonal matrices whose $i$-th entries are the first and half of the second derivatives of $g_i$ evaluated at the fixed point, respectively, and $()^2$ is the entry-wise square of the vector (see Supplement for details). By studying quadratic reservoirs and how they learn to manipulate their representations of chaotic manifolds, we will gain an intuition due to their analytic tractability, and generalizability across many activation functions $\bm{g}$ when driven within a range over which the quadratic expansion is accurate.

\section{Learning a translation operation by example}
\begin{figure}[h!]
	\centering
\textit{}	\includegraphics[width=6.5in]{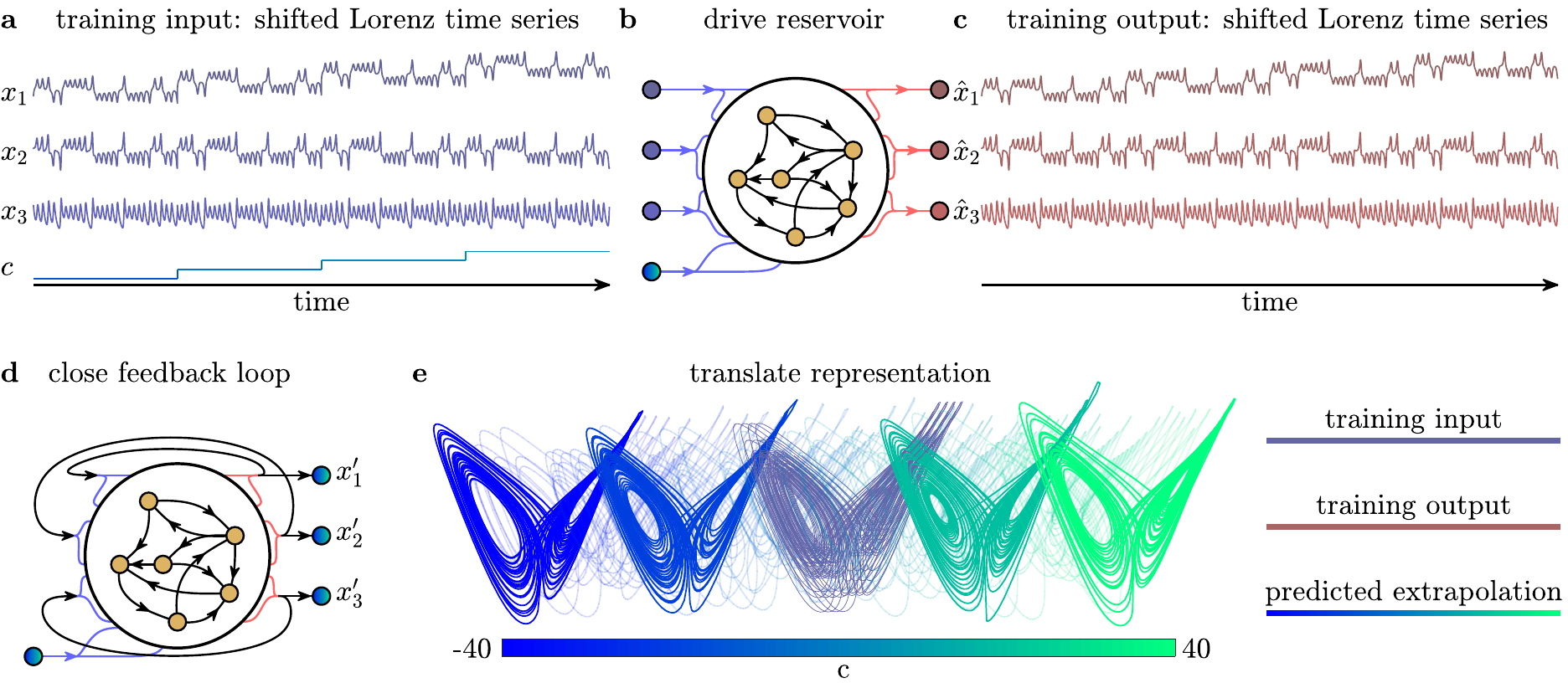}
	\caption{\textbf{Learning and extrapolating a translation operation through examples.} (\textbf{a}) Schematic of the time series of the Lorenz and control inputs, beginning with the original Lorenz time series $\bm{x}_0(t)$ at $\bm{c}=0$, followed by three equally spaced shifts in the $x_1$ direction and in the $\bm{c}$ parameter. (\textbf{b}) These inputs generate four reservoir time series $\bm{r}_c(t)$. (\textbf{c}) Next, weighted sums of the reservoir states are used to generate outputs $W\bm{r}_c(t) = \hat{\bm{x}}_c(t) \approx \bm{x}_c(t)$ that mimic the inputs. (\textbf{d}) The outputs $W\bm{r}(t)$ replace the inputs $\bm{x}(t)$ to create a reservoir with a closed feedback loop. (\textbf{e}) Over the course of a single simulation, the reservoir evolves autonomously about a Lorenz-shaped manifold, and translates this representation along $x_1$ by smoothly and continuously changing $\bm{c}$ as a real number over a range much larger than the training range.}
	\label{fig:f3}
\end{figure}

Reservoirs learn complex information through simple imitation: approximating the driving inputs using the reservoir states is enough to autonomously represent and evolve about a chaotic manifold. Here we show that this simple scheme is also enough to learn to translate the representation. We begin with a Lorenz time series $\bm{x}_0(t)$, and create shifted copies
\begin{align}
\label{eq:translation}
\bm{x}_c(t) = \bm{x}_0(t) + P\bm{c}.
\end{align}
For the purposes of demonstration, we consider a translation in the $x_1$ direction such that $P = [1; 0; 0]$ is a column vector, and $\bm{c} = 0,1,2,3$ is a scalar. We use these four time series to drive our reservoir according to Eq.~\ref{eq:reservoirc}, 
thereby generating four reservoir time series $\bm{r}_c(t)$. Numerically, $\bm{x}_c(t)$ and $\bm{r}_c(t)$ are matrices of dimension $3 \times T$ and $N \times T$ over $T$ time steps, which we concatenate along the time dimension into $\bm{x}(t) = [\bm{x}_0(t),\bm{x}_1(t),\bm{x}_2(t),\bm{x}_3(t)]$ and $\bm{r}(t) = [\bm{r}_0(t),\bm{r}_1(t),\bm{r}_2(t),\bm{r}_3(t)]$, respectively. Then, we compute output weights
\begin{align}
\label{eq:training}
W = \argmin_W\|W\bm{r}(t) - \bm{x}(t)\|_2,
\end{align}
such that our output $\hat{\bm{x}} = W\bm{r}(t)$ approximates our input $\bm{x}(t)$ (Fig.~\ref{fig:f3}a--c). Finally, we substitute the output as the input to yield the feedback system (Fig.~\ref{fig:f3}d)
\begin{align}
\label{eq:reservoirfbc}
\frac{1}{\gamma} \delta\dot{\bm{r}}' = -\delta\bm{r}' + U(R\delta\bm{r}' + C\bm{c}) + V(R\delta\bm{r}' + C\bm{c})^2,
\end{align}
where $R = A + BW$ (see Supplement for a discussion on $W\bm{r}_c(t) \approx W\delta\bm{r}_c(t)$).

As we evolve this autonomous reservoir while varying $\bm{c}$ to extreme values $-40 \leq \bm{c} \leq 40$ both inside and outside of the training values, it has learned to evolve about a Lorenz-shaped manifold that is translated based on the value of $\bm{c}$ (see Supplement for translations in all spatial directions). Hence, by training the network on shifted copies of the input time series, the reservoir has learned a translation operation on the attractor.

\section{Learning a linear transformation operation by example}
\begin{figure}[h!]
	\centering	\includegraphics[width=6.5in]{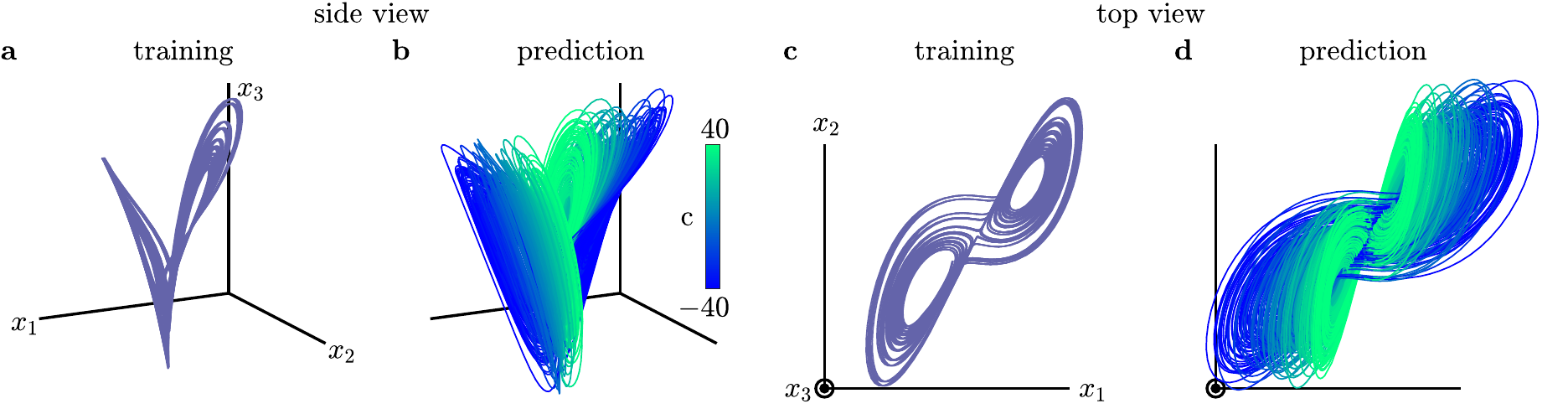}
	\caption{\textbf{Extrapolating a transformation operation through examples.} (\textbf{a}) 3-dimensional plot of the training data of the Lorenz time series that has been stretched along the $x_1$ direction at $c = 0,1,2,3$. (\textbf{b}) 3-dimensional plot of the feedback reservoir output that autonomously evolves about a Lorenz-shaped manifold that stretches dramatically based on varying $c$ from $-40$ to $40$. We also provide a top view of the (\textbf{c}) training data and (\textbf{d}) predicted output data.}
	\label{fig:f4}
\end{figure}

In addition to learning a translation operation that does not change the geometry of the representation, here we demonstrate that reservoirs can learn linear transformation using the exact same framework. Similarly, we begin with a Lorenz time series $\bm{x}_0(t)$ generated from Eq.~\ref{eq:lorenz}, and create linearly transformed copies of the time series such that
\begin{align}
\label{eq:transformation}
\bm{x}_c(t) = (I+cP)\bm{x}_0(t),
\end{align}
for $c = 0,1,2,3$, where $P$ is a matrix encoding a transformation (Fig.~\ref{fig:f4}a,c). Specifically, we perform a squeeze along $x_1$ by setting $[P]_{11} = -0.012$ and the remaining elements to 0. 

Exactly as before, we drive the reservoir according to Eq.~\ref{eq:reservoirc}, concatenate our input and reservoir time series into $\bm{x}(t)$ and $\bm{r}(t)$ to train the output weights $W$ according to Eq.~\ref{eq:training}, and feed the outputs back as inputs to yield the feedback system Eq.~\ref{eq:reservoirfbc}. This reservoir autonomously evolves about a Lorenz-shaped manifold that stretches based on the parameter $-40 \leq c \leq 40$ (Fig.~\ref{fig:f4}b,d) far outside of the parameters used in the training regime $c = 0,1,2,3$ (see Supplement for more examples). Hence, using the same framework, the reservoir has learned the linear transformation operation on the attractor manifold.

\section{Learning to infer a bifurcation by example}
For both translations and transformations, the reservoir learned a smooth change in its representation of the chaotic manifold. Here we demonstrate that a reservoir can infer, without actually ever having experienced, a much more dramatic change: a bifurcation. In the Lorenz attractor (Eq.~\ref{eq:lorenz} for $\rho > 1,\sigma = 10, \beta = 8/3$), there are two fixed points: one at the center of each wing, which undergo a subcritical Hopf bifurcation when $\rho = \rho^* \approx 24.7$ \cite{Strogatz1994}. When $\rho < \rho^*$, these two fixed points are stable. When $\rho > \rho^*$, the fixed points become unstable, yielding the characteristic wing-shaped flow. Here we demonstrate that a reservoir trained only on stable examples ($\rho < \rho^*$) can accurately predict the unstable flow ($\rho > \rho^*$).
\begin{figure}[h!]
	\centering	
	\includegraphics[width=6.5in]{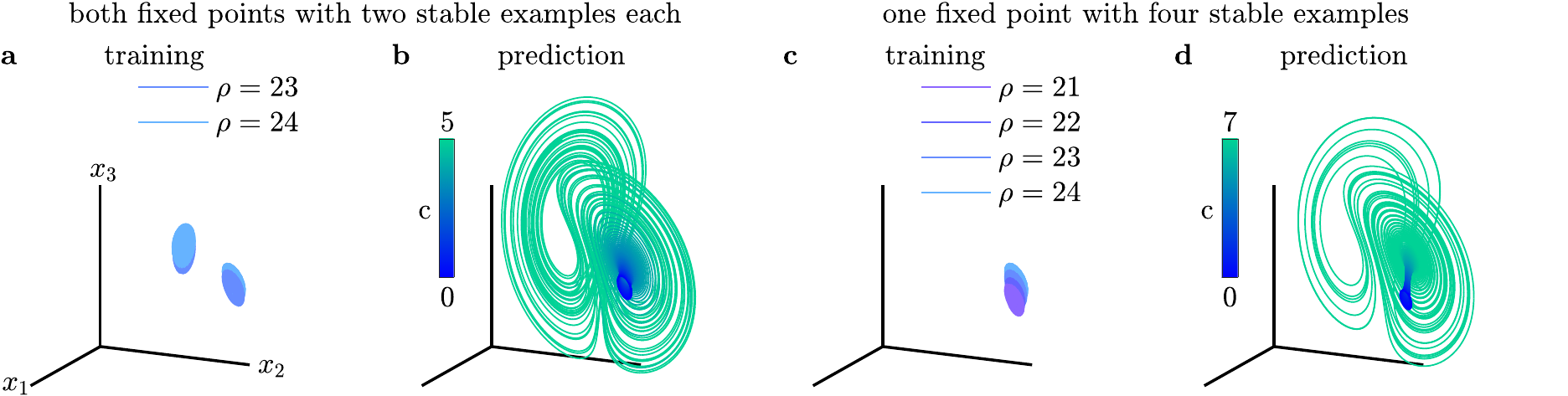}
	\caption{\textbf{Extrapolating the bifurcation of the Lorenz.} (\textbf{a}) Two training trajectories for each of the stable Lorenz fixed points at the wings, for $\rho=23$ with $c=0$ (blue) and for $\rho=24$ with $c=1$ (light blue). (\textbf{b}) The predicted trajectory of the feedback reservoir moves towards a stable fixed point for $c=0$, and bifurcates into a Lorenz-shaped manifold as $c$ is increased. (\textbf{c}) Four training examples for one of the stable Lorenz fixed points for $\rho = 21,22,23,24$ with $c = 0,1,2,3$. (\textbf{d}) The predicted trajectory moves towards a stable fixed point for $c=0$, and then bifurcates into a Lorenz-shaped manifold as $c$ is increased.}
	\label{fig:f5}
\end{figure}

For the two fixed points $\bm{a}$ and $\bm{b}$, we begin with four training trajectories: $\bm{x}^a_{23}(t)$ and $\bm{x}^b_{23}(t)$ that evolve stably towards the fixed points for $\rho = 23$, and $\bm{x}^a_{24}(t)$ and $\bm{x}^b_{24}(t)$ that evolve stably towards the fixed points for $\rho = 24$ (Fig.~\ref{fig:f5}a). We then drive the reservoir with $\bm{x}^a_{23}(t)$ and $\bm{x}^b_{23}(t)$ while setting $c=0$, and with $\bm{x}^a_{24}(t)$ and $\bm{x}^b_{24}(t)$ while setting $c=1$, and train the output weights. Finally, we evolve the feedback reservoir while changing $c$ from $0$ to $5$, and note that the trajectory bifurcates into a Lorenz-shaped manifold (Fig.~\ref{fig:f5}b). 

As a second demonstration, we begin with another set of four training trajectories: $\bm{x}^a_{21}(t),\dotsm,\bm{x}^a_{24}(t)$ that evolve stably towards only one fixed point for $\rho = 21,\dotsm,24$ (Fig.~\ref{fig:f5}c). We then drive the reservoir with $\bm{x}^a_{21}(t),\dotsm,\bm{x}^a_{24}(t)$ while setting $c = 0,\dotsm,3$, and train the output weights. Finally, we evolve the feedback reservoir while changing $c$ from $0$ to $7$, and note that the trajectory again bifurcates into a Lorenz-shaped manifold (Fig.~\ref{fig:f5}d). Hence, after only observing a few stable trajectories before the bifurcation ($\rho < \rho^*$), the reservoir accurately extrapolates the geometry of the Lorenz trajectory after the bifurcation ($\rho > \rho^*$).

\section{Mechanism of how operations are learned}
Now that we have taught reservoirs to manipulate chaotic manifolds, we seek to understand the mechanism. We begin with some intuition by expanding the feedback dynamics
\begin{align*}
\frac{1}{\gamma} \delta\dot{\bm{r}}' = ([U + 
\underbrace{2V~\mathrm{diag}(C\bm{c})}_{\mathrm{stretch}}]R - I)\delta\bm{r}' + \underbrace{UC\bm{c}}_{\mathrm{shift}} + V(R\delta\bm{r}')^2+ \underbrace{V(C\bm{c})^2}_{\mathrm{small}},
\end{align*}
and notice that the control parameter can scale the shape of the reservoir's internal dynamics (stretch), and add a constant driving input (shift). For small changes in $\bm{c}$, the quadratic term $C\bm{c}$ is negligible. To formalize this intuition, we consider the time series $\bm{r}'(t) = \bm{r}'_{c=0}(t)$ generated by evolving the autonomous reservoir according to Eq.~\ref{eq:reservoirfbc} at $\bm{c} = \bm{0}$. Next, we take the total differential of Eq.~\ref{eq:reservoirfbc} evaluated at $\bm{r}'(t)$ and $\bm{c}=\bm{0}$ to yield
\begin{align}
\label{eq:differential}
(I-KA)d\bm{r}' + \frac{1}{\gamma}d\dot{\bm{r}}' = K(BWd\bm{r}' + Cd\bm{c}),
\end{align}
where $K = U + 2V~\mathrm{diag}(R\delta\bm{r}'(t))$. Our goal is to write the change in the reservoir state $d\bm{r}'(t)$ that is induced by changing the control parameter by an infinitesimal amount $d\bm{c}$. 

\begin{figure}[h!]	\includegraphics[width=6.5in]{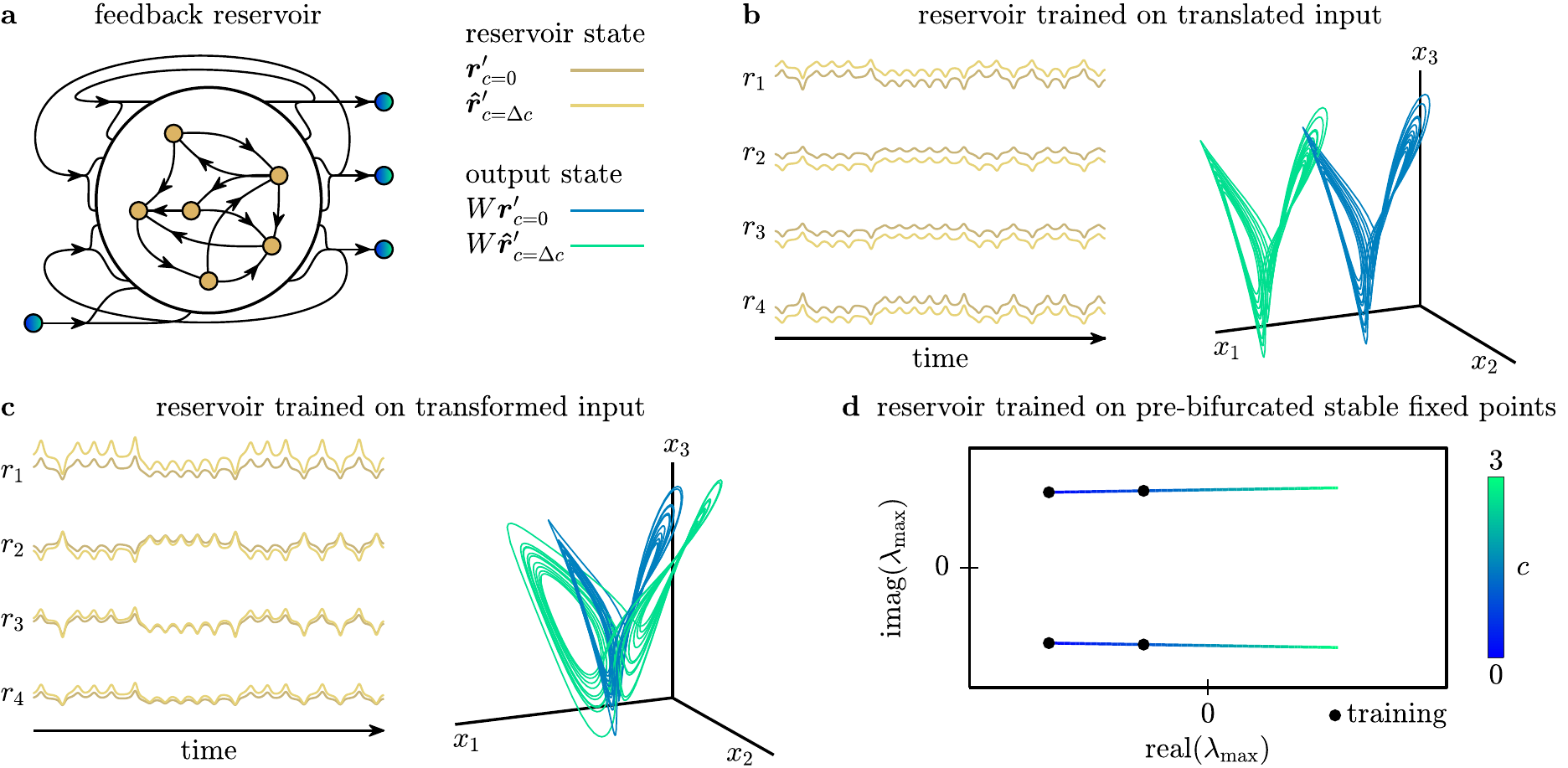}
	\caption{\textbf{Changing the control parameter changes the reservoir dynamics to manipulate representations.} (\textbf{a}) Schematic of a reservoir with feedback connections after the output weights $W$ have been trained. (\textbf{b}) Reservoir time series generated by evolving the autonomous reservoir with the original Lorenz input with $\bm{c} = 0$ (dark gold). We also show the predicted time series from solving Eq.~\ref{eq:predict_translate} after training on translated examples and setting $d\bm{c} = \Delta \bm{c} = 20$ (light gold). The output projections of the two time series are shown in blue and green, respectively. (\textbf{c}) The original and predicted reservoir states and their output projections for $\Delta c = -40$ after training on transformed Lorenz inputs by solving Eq.~\ref{eq:predict_transform}. (\textbf{d}) Plot of the real and imaginary components of the two most unstable eigenvalues of the autonomous reservoir trained on two stable Lorenz trajectories (Fig.~\ref{fig:f5}a,b). The reservoir is linearized about its fixed point according to Eq.~\ref{eq:reservoirfblin} as $c$ is slowly changed.}
	\label{fig:f6}
\end{figure}

When learning translations, the output weights are trained such that $W\bm{r}_c(t) \approx \bm{x}_c(t) = \bm{x}(t) + P\bm{c}$. For sufficiently nearby training examples (small $P, \bm{c}$), we also implicitly approximate the differential relation $Wd\bm{r}(t) \approx Pd\bm{c}$. Additionally, if the feedback reservoir stabilizes these examples, then $Wd\bm{r}'(t) \approx Pd\bm{c}$. Substituting this relation into Eq.~\ref{eq:differential} yields
\begin{align*}
(I-KA)d\bm{r}' + \frac{1}{\gamma}d\dot{\bm{r}}' \approx K(BP + C)d\bm{c}.
\end{align*}
If we fix $d\bm{c}$, we have $2N$ variables, $d\bm{r}'$ and $d\dot{\bm{r}}'$, but only $N$ equations. By taking the time derivative of the differential relation, we generate another $N$ variables and $N$ equations. Continuing to take time derivatives yields the following system of equations
\begin{align*}
\begin{bmatrix}
H_0 & H_{-1} & 0 & \dotsm\\
H_1 & H_0    & H_{-1} & \dotsm\\
H_2 & 2H_1   & H_0 & \dotsm\\
\vdots & \vdots& \vdots & \ddots
\end{bmatrix}
\begin{bmatrix}
d\bm{r}'\\
d\dot{\bm{r}}'\\
d\ddot{\bm{r}}'\\
\vdots
\end{bmatrix}
\approx
\begin{bmatrix}
K\\
\dot{K}\\
\ddot{K}\\
\vdots
\end{bmatrix}
(BP + C)d\bm{c},
\end{align*}
where $H_{-1} = \frac{1}{\gamma}I$, $H_0 = I-KA$, and $H_i = -K^{(i)}A$ is the $i$-th time-derivative of $KA$. This matrix is a block-Hessenberg matrix, with an analytic solution \cite{Sowik2018} for the first term $d\bm{r}'$. We truncate this solution (see Supplement) to explicitly relate $d\bm{c}$ to $d\bm{r}'$ as follows:
\begin{align}
\label{eq:predict_translate}
d\bm{r}' \approx 
-\begin{bmatrix}
\gamma H_0^2 - H_1
\end{bmatrix}^{-1} 
\begin{bmatrix}
-\gamma H_0 & ~I
\end{bmatrix}
\begin{bmatrix}
K\\
\dot{K}
\end{bmatrix}
(BP + C)d\bm{c}.
\end{align}
As a demonstration, we pick a finite $\Delta\bm{c} = 20$, and plot the original and predicted change in the reservoir states, and their outputs in spatial coordinates (Fig.~\ref{fig:f6}b). Hence, using only the feedback dynamics Eq.~\ref{eq:reservoirfbc} and sufficiently nearby training examples, changing $\bm{c}$ causes changes in the reservoir states from Eq.~\ref{eq:predict_translate} that map to a translation.

The same approach can be used for linear transformations, where the output weights are trained such that $W\bm{r}_c(t) \approx \bm{x}_c(t) = (I+cP)\bm{x}(t)$. For sufficiently nearby training examples, we implicitly approximate the differential relation $Wd\bm{r}(t) \approx P\bm{x}(t)dc \approx PW\bm{r}(t)dc$, which if properly stabilized, yields $Wd\bm{r}'(t) \approx PW\bm{r}'(t)$. Performing the same time derivatives and solution truncation as in the translation, we get the following relation between $dc$ and $d\bm{r}'$:
\begin{align}
\label{eq:predict_transform}
d\bm{r}' \approx 
-\begin{bmatrix}
\gamma H_0^2 - H_1
\end{bmatrix}^{-1} 
\begin{bmatrix}
-\gamma H_0 & ~I
\end{bmatrix}
\begin{bmatrix}
K(BPW\bm{r}' + C)\\
\dot{K}(BPW\bm{r}'+C) + KBPW\dot{\bm{r}}'
\end{bmatrix}
dc.
\end{align}
As another demonstration, we set $\Delta c = -40$, and plot the original and predicted change in the reservoir states, and their outputs (Fig.~\ref{fig:f6}c).

Finally, to understand how the reservoir is able to infer a bifurcation, we demonstrate that it learns a smooth translation of eigenvalues. Specifically, at $\rho^*$, the fixed points at the wings of the Lorenz system undergo a Hopf bifurcation, whereby the real component of complex conjugate eigenvalues goes from negative to positive. To track the eigenvalues of the autonomous reservoir, we linearize Eq.~\ref{eq:reservoirfbc} about a fixed point $\delta\bm{r}^*$ such that
\begin{align}
\label{eq:reservoirfblin}
\frac{1}{\gamma}\delta\dot{\bm{r}}' \approx [-I + UR + 2V\mathrm{diag}(R\delta\bm{r}^* + C\bm{c})R](\delta\bm{r}'-\delta\bm{r}^*).
\end{align}
Then, using the output weights trained only on stable Lorenz trajectories (at $c = 0, \rho = 23$ and $c = 1, \rho = 24$; Fig.~\ref{fig:f5}a,b), we track the autonomous reservoir's two most unstable eigenvalues (largest real component) at the fixed point as we vary the control parameter from $c = 0$ to $c = 3$. We find that these eigenvalues are complex conjugates whose real components go from negative to positive (Fig.~\ref{fig:f6}d). Hence, we demonstrate that not only can reservoirs learn smooth translations and transformations by mapping $d\bm{c}$ to $d\bm{r}'$, but they can also perform bifurcations by learning smooth changes in their eigenvalues.

\section{Simultaneous learning of multiple operations}

\begin{figure}[h!]
	\centering	
	\includegraphics[width=6.5in]{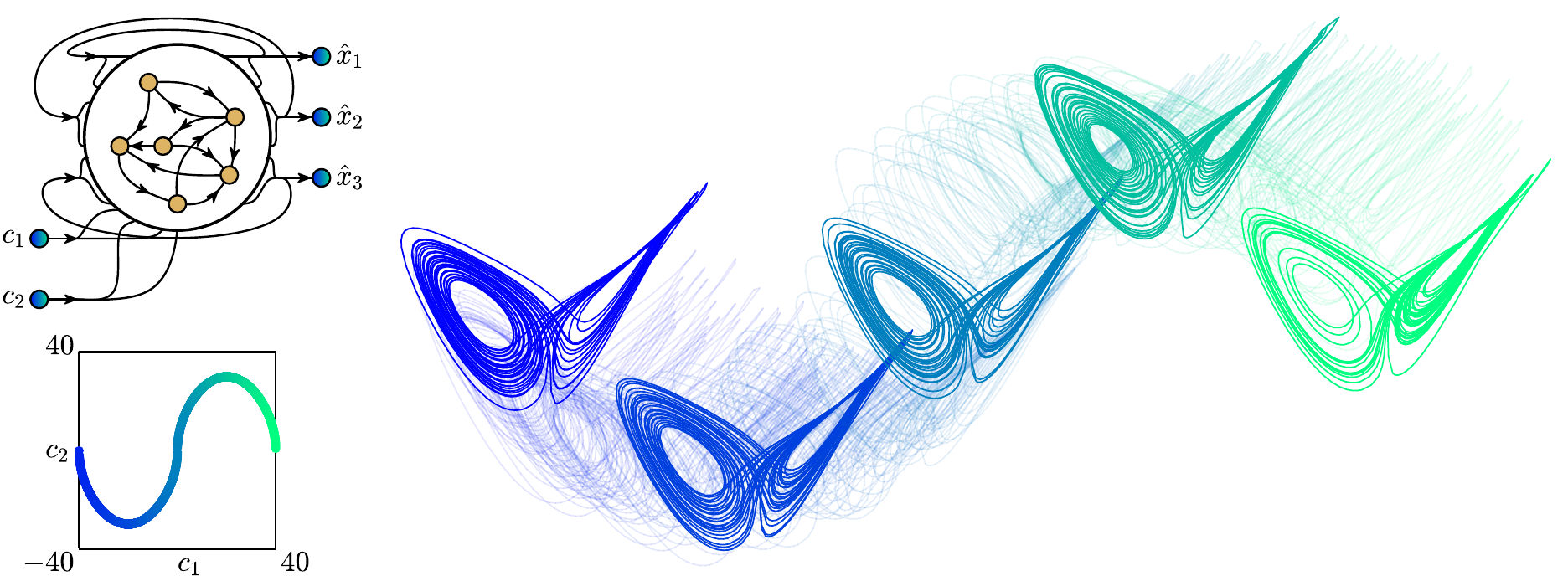}
	\caption{\textbf{Flight of the Lorenz.} A reservoir trained on translated inputs along the $x_1$ and $x_3$ directions evolves autonomously along a Lorenz-shaped chaotic manifold. We can change the $x_1$ and $x_3$ position of its representation by changing control parameters $c_1$ and $c_2$, respectively.}
	\label{fig:f7}
\end{figure}

To close, here we demonstrate that reservoirs can easily learn multiple computations by changing multiple control inputs. We train a translation in the $x_1$ direction with control parameter $c_1$, and a translation in the $x_3$ direction with control parameter $c_2$. As before, we begin with a Lorenz time series $\bm{x}_{0,0}(t)$ generated from Eq.~\ref{eq:lorenz}, and created shifted copies
\begin{align*}
\bm{x}_{c_1,c_2}(t) = \bm{x}_{0,0}(t) + c_1\bm{a}_1 + c_2\bm{a}_2,
\end{align*}
where $\bm{a}_1 = [1;0;0]$ corresponds to an $x_1$ shift, and $\bm{a}_2 = [0;0;1]$ corresponds to an $x_3$ shift. We generate 10 shifted inputs, with one unshifted attractor ($c_1 = 0, c_2 = 0$), three shifts in the $x_1$ direction ($c_1 = 1,2,3, c_2 = 0$), three shifts in the $x_3$ direction ($c_1 = 0, c_2 = 1,2,3$), and three shifts in both directions ($c_1 = 1,2,3, c_2 = 1,2,3$). We use these shifted copies along with their corresponding control inputs to drive our reservoir and produce 10 reservoir time series $\bm{r}_{c_1,c_2}(t)$. Then, we concatenate these 10 time series into $\bm{x}(t)$ and $\bm{r}(t)$ to train output weights $W$ according to Eq.~\ref{eq:training}, and perform the feedback according to Eq.~\ref{eq:reservoirfbc} where $\bm{c} = [c_1;~c_2]$ is a vector. By changing parameters $c_1$ and $c_2$, the reservoir evolves about a Lorenz-shaped manifold that is shifted in the $x_1$ and $x_3$ directions (Fig.~\ref{fig:f7}).

\section{Discussion}
In this paper, we teach an RNN to evolve about a Lorenz-shaped manifold, and to control its evolution about a translated, transformed, and bifurcated continua of such manifolds. Our approach contributes to prior work on artificial neural networks in three significant ways \cite{Seung1998learning,Wu2016,Jaeger2010,Sussillo2009}. First, we provide a means by which a neural system can learn continuous interpolated and extrapolated modifications, along with discontinuous bifurcations, of its own representation solely through examples. Second, the learned manifolds are spatially and temporally complex, allowing for potential extensions to learning modifications of time series data such as speech or music with a structured yet unpredictable evolution. Third, we use a randomly generated and arbitrarily connected network that does not need to be artificially engineered to preserve invariance or manipulate information \cite{Wu2016}. 

One of the main limitations of this work is the lack of a clear mechanism of how the network connectivity ultimately stabilizes the chaotic manifold. Much progress has been made in tackling this limitation, both by exercising theoretical concepts of generalized synchronization \cite{Rulkov1995}, and by developing tools for controlling chaos \cite{Ott1990}. However, there is insufficient knowledge to guarantee that a set of training and reservoir parameters will always successfully teach the desired computation. Similarly, we are unable to specify exactly how far to space the training examples for the feedback reservoir to successfully learn the linear relationships between the differential of the reservoir states and the control parameter.

A particularly promising area for future work is related to the simple quadratic form of the reservoir. Because all of these results are obtained by driving our reservoir in the quadratic regime, the same results should hold for common neural mass models, such as the Wilson-Cowan model \cite{Wilson1972}. Hence, these results may provide a unifying framework for learning and computing in dynamical neural models. Additionally, these results provide a basis for exploring more complex computations, such as inferring bifurcations in experimental data, and testing the reservoir's ``imagination'' in reconstructing more complex chaotic manifolds using incomplete data. Finally, and perhaps most astonishingly, the reservoir's ability to accurately reconstruct the full nonlinear geometry of the bifurcated Lorenz manifold after only observing pre-bifurcation data implies that it is not only imitating examples, but actually inferring higher-order nonlinear structure. This work therefore provides a starting point for exploring exactly how higher-order structure is learned by neural systems.

\section{Acknowledgments}
We are thankful for the insightful feedback and comments from Harang Ju and Keith Wiley. We gratefully acknowledge support from the John D. and Catherine T. MacArthur Foundation, the Alfred P. Sloan Foundation, the ISI Foundation, the Paul Allen Foundation, the Army Research Laboratory (No. W911NF-10-2-0022), the Army Research Office (Nos. Bassett-W911NF-14-1-0679, Grafton-W911NF-16-1-0474, and DCIST-W911NF-17-2-0181), the Office of Naval Research (ONR), the National Institute of Mental Health (Nos. 2-R01-DC-009209-11, R01-MH112847, R01-MH107235, and R21-M MH-106799), the National Institute of Child Health and Human Development (No. 1R01HD086888-01), National Institute of Neurological Disorders and Stroke (No. R01 NS099348), and the National Science Foundation (NSF) (Nos. DGE-1321851, BCS-1441502, BCS-1430087, NSF PHY-1554488, and BCS-1631550). The content is solely the responsibility of the authors and does not necessarily represent the official views of any of the funding agencies.

\section{Citation Diversity Statement}
Recent work in several fields of science has identified a bias in citation practices such that papers from women and other minorities are under-cited relative to other papers in the field \cite{Dworkin2020}. Here we sought to proactively consider choosing references that reflect the  diversity of our field in thought, form of contribution, gender, and other factors. We classified gender based on the first names of the first and last authors, with possible combinations including male/male, male/female, female/male, and female/female. We regret that our current methodology is limited to consideration of gender as a binary variable. Excluding self-citations to the first and senior authors of our present paper, the references contain 50\% male/male, 23.5\% male/female, 11.8\% female/male, and 14.7\% female/female categorizations. We look forward to future work that will help us to better understand how to support equitable practices in science.

\newpage
\section{References}
\bibliography{references}

\end{document}


\bibliographystyle{naturemag}
\title{Supplement to ``Teaching Recurrent Neural Networks to Modify Chaotic Memories by Example''}
\author{Jason Z. Kim}
\affiliation{Department of Bioengineering, University of Pennsylvania, Philadelphia, PA, 19104}
\author{Zhixin Lu}
\affiliation{Department of Bioengineering, University of Pennsylvania, Philadelphia, PA, 19104 }
\author{Erfan Nozari}
\affiliation{Department of Electrical and Systems Engineering, University of Pennsylvania, Philadelphia, PA, 19104 }
\author{George J. Pappas}
\affiliation{Department of Electrical and Systems Engineering, University of Pennsylvania, Philadelphia, PA, 19104}
\author{Danielle S. Bassett}
\affiliation{Departments of Bioengineering, Physics \& Astronomy, Electrical \& Systems Engineering, Neurology, and Psychiatry, University of Pennsylvania, Philadelphia, PA, 19104}
\affiliation{Santa Fe Institute, Santa Fe, NM 87501}
\affiliation{To whom correspondence should be addressed: dsb@seas.upenn.edu}
\date{\today}
\maketitle
\tableofcontents
\newpage

\section{Methods}
In this section, we describe additional details about the methods and simulations used in the main text. We begin with a more thorough overview of reservoir dynamics and their derivation, followed by specific details of the numerical simulations.

\subsection{Reservoir dynamics}
The reservoir computing framework is a general scheme by which a nonlinear dynamical system (the reservoir) is driven by some input, and a simple linear readout of the reservoir states is trained \cite{Lukosevicius2012}. The reservoir consists of $N$ neural units, where each unit $i$ has a real-valued level of activity over time, $r_i(t)$. We collect this activity into an $N$-dimensional column vector
\begin{align*}
\bm{r}(t) = 
\begin{bmatrix}
r_1(t)\\ r_2(t)\\ \vdots \\ r_N(t)
\end{bmatrix},
\end{align*}
that we refer to as the \emph{reservoir state}. These reservoir states are driven by some input time series of $M$ inputs $x_1(t), x_2(t), \dotsm, x_M(t)$, that we collect into the input vector
\begin{align*}
\bm{x}(t) = \begin{bmatrix}
x_1(t)\\ x_2(t)\\ \vdots\\ x_M(t)
\end{bmatrix}.
\end{align*}
In our framework, we add a set of $K$ control inputs $c_1,\dotsm,c_K$ that we collect into the control vector
\begin{align*}
\bm{c} = 
\begin{bmatrix}
c_1\\ c_2\\ \vdots\\ c_K
\end{bmatrix}.
\end{align*}

For continuous time systems ($t \in \real_{\geq 0}$), a typical equation for the time-evolution of a reservoir consists of a nonlinear (usually sigmoidal \cite{Lukosevicius2012}) transformation $\bm{g}$ on a linear sum of all inputs and states written as
\begin{align*}
\frac{1}{\gamma} \dot{\bm{r}}(t)  = -\bm{r}(t) + \bm{g}(A\bm{r}(t) + B\bm{x}(t) + C\bm{c} + \bm{d}),
\end{align*}
where $\dot{\bm{r}}(t)$ represents the time derivative \cite{Jaeger2010,Sussillo2009}, $A$ is a real-valued matrix of dimension $N \times N$, $B$ is a real-valued matrix of dimension $N \times M$, $C$ is a real-valued matrix of dimension $N \times K$, and $\bm{d}$ is a constant bias vector of dimension $N \times 1$. We can write the dynamics for each reservoir state, $r_i(t)$, as
\begin{align*}
\frac{1}{\gamma}\dot{r}_i(t) = -r_i(t) + g_i\left(\sum_{n=1}^N A_{in}r_n(t) + \sum_{m=1}^M B_{im}x_m(t) + \sum_{k=1}^K C_{ik}c_k + d_i\right).
\end{align*}
If we write $A_{i*}, B_{i*}$, and $C_{i*}$ as the $i$-th row of matrices $A, B$, and $C$, respectively, we can write this equation more concisely as
\begin{align}
\label{eq:idynamics}
\frac{1}{\gamma}\dot{r}_i(t) = -r_i(t) + g_i\left(A_{i*}\bm{r}(t) + B_{i*}\bm{x}(t) + C_{i*}\bm{c} + d_i\right).
\end{align}

We begin by observing that the reservoir states are evolved according to some predetermined input $\bm{x}(t)$ and control input $\bm{c}$ to generate the reservoir state time series $\bm{r}(t)$. Next, linear combinations of the reservoir state are taken to approximate the input $\bm{x}(t)$ by minimizing the matrix 2-norm of the difference in the numerical time series (see Sec. \ref{sec:training})
\begin{align*}
\|W\bm{r}(t)  - \bm{x}(t)\|_2,
\end{align*}
where $W$ is the real valued matrix of dimension $M \times N$ that is trained. After training, we perform feedback by replacing the inputs $\bm{x}(t)$ with the trained outputs $W\bm{r}(t)$ to yield the feedback dynamics
\begin{align*}
\frac{1}{\gamma} \dot{\bm{r}}'(t)  = -\bm{r}'(t) + \bm{g}(A\bm{r}'(t) + BW\bm{r}'(t) + C\bm{c} + \bm{d}),
\end{align*}
and by factoring the term $R = A+BW$, we obtain
\begin{align*}
\frac{1}{\gamma} \dot{\bm{r}}'(t)  = -\bm{r}'(t) + \bm{g}(R\bm{r}'(t) + C\bm{c} + \bm{d}).
\end{align*}
This feedback equation is written element-wise as
\begin{align*}
\frac{1}{\gamma}\dot{r}'_i(t) = -r'_i(t) + g_i\left(\sum_{n=1}^N R_{in}r'_n(t) + \sum_{k=1}^K C_{ik}c_k + d_i\right).
\end{align*}
We note that $R$ is an $N \times N$ matrix.
\newpage

\subsection{Derivation of quadratic reservoir}
While the reservoir computing framework often uses a specific nonlinear sigmoid function for $\bm{g}$, our goal is to find principles of learning relevant to a broad range of functional forms of $\bm{g}$. To achieve this goal, we study the reservoir in the weakly nonlinear regime. By deriving insights in this regime, we aim to make statements about dynamical systems with many different forms of $\bm{g}$ as long they are driven in the same regime. By weakly nonlinear, we mean the \emph{quadratic} regime, where the reservoir evolves nearby some constant stable fixed point attractor $\bm{r}^*$. This regime is explicitly encoded by taking the second-order Taylor series expansion of the dynamics about a steady state in all of the inputs $\bm{r}^*, \bm{x}^*, \bm{c}^*$. For notational convenience, we write $\delta\bm{r} = \bm{r}-\bm{r}^*$, $\delta\bm{x} = \bm{x}-\bm{x}^*$, and $\delta\bm{c} = \bm{c}-\bm{c}^*$, and also omit the notation of time dependence, $(t)$. If we write the full dynamics from Eq.~\ref{eq:idynamics} as
\begin{align*}
\frac{1}{\gamma}\dot{r}_i = f_i(\bm{r},\bm{x},\bm{c}) = -r_i + g_i\left(A_{i*}\bm{r} + B_{i*}\bm{x} + C_{i*}\bm{c} + d_i\right),
\end{align*}
then the Taylor series expansion to second order contains terms
\begin{align*}
T_{i,0} &= f_i|_{\bm{r}=\bm{r}^*,\bm{x}=\bm{x}^*,\bm{c}=\bm{c}^*}\\
T_{i,1} &= \nabla_{\bm{r},\bm{x},\bm{c}} f_i|_{\bm{r}=\bm{r}^*,\bm{x}=\bm{x}^*,\bm{c}=\bm{c}^*}\\
T_{i,2} &= \nabla^2_{\bm{r},\bm{x},\bm{c}} f_i|_{\bm{r}=\bm{r}^*,\bm{x}=\bm{x}^*,\bm{c}=\bm{c}^*},
\end{align*}
where $\nabla$ is the gradient operator with respect to the subscripted variables
\begin{align*}
\nabla_{\bm{r},\bm{x},\bm{c}} 
&= 
\begin{bmatrix}
\nabla_{\bm{r}}, & \nabla_{\bm{x}}, & \nabla_{\bm{c}}
\end{bmatrix}\\
&=
\begin{bmatrix}
\frac{\partial}{\partial r_1},  & \dotsm, & \frac{\partial}{\partial r_N}, & 
\frac{\partial}{\partial x_1}, & \dotsm, &
\frac{\partial}{\partial x_M}, &
\frac{\partial}{\partial c_1}, & \dotsm, &
\frac{\partial}{\partial c_K}
\end{bmatrix},
\end{align*}
yielding a vector of partial derivatives, and $\nabla^2$ yields a matrix of all pairwise second partial derivatives. Then the quadratic dynamics become
\begin{align*}
\frac{1}{\gamma} \delta\dot{r}_i \approx T_{i,0} + T_{i,1} \begin{bmatrix}
\delta\bm{r}\\ \delta\bm{x}\\ \delta\bm{c}
\end{bmatrix}
+
\frac{1}{2}
\begin{bmatrix}
\delta\bm{r}^\top & \delta\bm{x}^\top & \delta\bm{c}^\top
\end{bmatrix}
T_{i,2} \begin{bmatrix}
\delta\bm{r}\\ \delta\bm{x}\\ \delta\bm{c}
\end{bmatrix}.
\end{align*}

As we are evaluating the dynamics about a fixed point, the reservoir does not change its state at this point, such that $f_i|_{\bm{r}=\bm{r}^*,\bm{x}=\bm{x}^*,\bm{c}=\bm{c}^*} = 0$. Hence the term
\begin{align*}
T_{i,0} = 0
\end{align*}
vanishes. The next term, $T_{i,1}$, is a linear approximation of the dynamics $f_i$
\begin{align*}
T_{i,1} 
&= \nabla_{\bm{r},\bm{x},\bm{c}}(-r_i) + \nabla_{\bm{r},\bm{x},\bm{c}} g_i\left(A_{i*}\bm{r} + B_{i*}\bm{x} + C_{i*}\bm{c} + d_i\right)\\
&= \begin{bmatrix} 0, & \dotsm, & 0, & -1, & 0, & \dotsm, & 0 \end{bmatrix} + u_i \begin{bmatrix}
A_{i*}, & B_{i*}, & C_{i*}
\end{bmatrix},
\end{align*}
where $u_i = \left. \frac{\partial g_i}{\partial (\bm{r}, \bm{x}, \bm{c})} \right|_{\bm{r}^*,\bm{x}^*,\bm{c}^*}$ is the evaluation of the first derivative of $g_i$ at the fixed point, and the subsequent term $T_{i,2}$ approximates the quadratic curvature of $f_i$ as follows:
\begin{align*}
T_{i,2} 
&= \nabla^2_{\bm{r},\bm{x},\bm{c}} f_i|_{\bm{r}=\bm{r}^*,\bm{x}=\bm{x}^*,\bm{c}=\bm{c}^*}\\
&= 2v_i
\begin{bmatrix}
A_{i*}^\top\\
B_{i*}^\top\\
C_{i*}^\top
\end{bmatrix}
\begin{bmatrix}
A_{i*}, & B_{i*}, & C_{i*}
\end{bmatrix},
\end{align*}
where $v_i =  \frac{1}{2}\left. \frac{\partial^2 g_i}{\partial (\bm{r}, \bm{x}, \bm{c})^2} \right|_{\bm{r}^*,\bm{x}^*,\bm{c}^*}$ is half of the evaluation of the second derivative of $g_i$ at the fixed point. 
Substituting these values back into the quadratic dynamics, we obtain
\begin{align*}
\frac{1}{\gamma} \delta\dot{r}_i \approx  
-\delta r_i
+
u_i \begin{bmatrix}
A_{i*}, & B_{i*}, & C_{i*}
\end{bmatrix} 
\begin{bmatrix}
\delta\bm{r}\\ \delta\bm{x}\\ \delta\bm{c}
\end{bmatrix} +
v_i
\begin{bmatrix}
\delta\bm{r}^\top & \delta\bm{x}^\top & \delta\bm{c}^\top
\end{bmatrix}
\begin{bmatrix}
A_{i*}^\top\\
B_{i*}^\top\\
C_{i*}^\top
\end{bmatrix}
\begin{bmatrix}
A_{i*}, & B_{i*}, & C_{i*}
\end{bmatrix} 
\begin{bmatrix}
\delta\bm{r}\\ \delta\bm{x}\\ \delta\bm{c}
\end{bmatrix},
\end{align*}
and we notice that
\begin{align*}
\begin{bmatrix}
\delta\bm{r}^\top & \delta\bm{x}^\top & \delta\bm{c}^\top
\end{bmatrix}
\begin{bmatrix}
A_{i*}^\top\\
B_{i*}^\top\\
C_{i*}^\top
\end{bmatrix}
=
\begin{bmatrix}
A_{i*}, & B_{i*}, & C_{i*}
\end{bmatrix} 
\begin{bmatrix}
\delta\bm{r}\\ \delta\bm{x}\\ \delta\bm{c}
\end{bmatrix},
\end{align*}
to yield
\begin{align*}
\frac{1}{\gamma} \delta\dot{r}_i \approx 
-\delta r_i
+
u_i
\begin{bmatrix}
A_{i*}, & B_{i*}, & C_{i*}
\end{bmatrix}
\begin{bmatrix}
\delta\bm{r}\\ \delta\bm{x}\\ \delta\bm{c}
\end{bmatrix}
+
v_i
\left(
\begin{bmatrix}
A_{i*}, & B_{i*}, & C_{i*}
\end{bmatrix} 
\begin{bmatrix}
\delta\bm{r}\\ \delta\bm{x}\\ \delta\bm{c}
\end{bmatrix}
\right)^2,
\end{align*}
which can be rewritten as
\begin{align*}
\frac{1}{\gamma}\delta\dot{r}_i \approx -\delta r_i + u_i (A_{i*}\delta\bm{r} + B_{i*}\delta\bm{x} + C_{i*}\delta\bm{c}) + v_i (A_{i*}\delta\bm{r} + B_{i*}\delta\bm{x} + C_{i*}\delta\bm{c})^2.
\end{align*}
Compiling the dynamics of all reservoir nodes $\bm{r}$, we write the compact vector form of the dynamics as
\begin{align*}
\frac{1}{\gamma}\delta\dot{\bm{r}} = -\delta\bm{r} + U(A\delta\bm{r} + B\delta\bm{x} + C\delta\bm{c}) + V(A\delta\bm{r} + B\delta\bm{x} + C\delta\bm{c})^2,
\end{align*}
where $U$ and $V$ are diagonal matrices where the $i$-th elements are $u_i$ and $v_i$ obtained by evaluating the first and second derivatives of $g_i$, respectively. To avoid making any assumptions about the operating point of the input states or control inputs, we linearize about $\bm{x}^* = \bm{0}$ and $\bm{c}^* = \bm{0}$. Further, we notice that $\delta\dot{\bm{r}} = \frac{d}{dt} (\bm{r} - \bm{r}^*) = \dot{\bm{r}}$. Hence, the above vector form of the equation becomes
\begin{align}
\label{eq:quadratic}
\frac{1}{\gamma}\delta \dot{\bm{r}} = -\delta \bm{r} + U(A\delta\bm{r} + B\bm{x} + C\bm{c}) + V(A\delta\bm{r} + B\bm{x} + C\bm{c})^2.
\end{align}

Importantly, we note that the elements of $U$ and $V$ are not allowed to arbitrarily be any value. Instead, their values depend on the first and second derivatives of the activation function $\bm{g}$ evaluated at the fixed points $\bm{r}^*,\bm{x}^*$, and $\bm{c}^*$. For the sake of remaining relevant to the existing literature that frequently uses the hyperbolic tangent function $tanh$ \cite{Lukosevicius2012}, we will restrict the values of $U$ and $V$. First, we note that at the fixed point, we have
\begin{align*}
\bm{0} = -\bm{r}^* + \bm{g}(A\bm{r}^* + B\bm{x}^* + C\bm{c}^*).
\end{align*}
Recall that we evaluate our functions at $\bm{x}^* = \bm{0}$ and $\bm{c}^* = \bm{0}$, such that $\bm{g}(A\bm{r}^*) = \bm{r}^*$, regardless of the form of $\bm{g}$. Next, we consider the specific activation function
\begin{align*}
\bm{g}(\bm{r},\bm{x},\bm{c}) = \tanh(A\bm{r} + B\bm{x} + C\bm{c} + \bm{d}),
\end{align*}
and evaluate its first derivative at $\bm{r}^*, \bm{x}^* = \bm{0}, \bm{c}^* = \bm{0}$ to yield
\begin{align}
\label{eq:U}
\mathrm{d} \bm{g}_{\bm{r}^*,\bm{x}^*=\bm{0},\bm{c}^*=\bm{0}} = \bm{1}-\tanh(A\bm{r}^*)^2 = \bm{1}-\bm{r}^{*2},
\end{align}
where $\bm{1}$ is an $N$-dimensional vector of ones, and the square notation of the vector implies an element-wise square. Hence, $U$ is a diagonal matrix where the $i$-th entry is $1 - r_i^{*2}$. Finally, we take the second derivative 
\begin{align}
\label{eq:V}
\mathrm{d}^2 \bm{g}_{\bm{r}^*,\bm{x}^*=\bm{0},\bm{c}^*=\bm{0}} = -2(\tanh(A\bm{r}^*) - \tanh(A\bm{r}^*)^3) = 2(\bm{r}^{*3} - \bm{r}^*),
\end{align}
such that the $i$-th diagonal element of $V$ is given by $r_i^{*3} - r_i^*$. Hence, for the quadratic approximation of $tanh$, it is sufficient to specify a fixed point $\bm{r}^*$ to fully determine the matrices $U$ and $V$.

\newpage

\subsection{Simulation parameters}
For all simulations in the main text of the paper, we used the following parameter choices:

\subsubsection{Global parameters}
\begin{itemize}
	\item simulation time step: $\Delta t = 0.001$.
\end{itemize}

\subsubsection{Lorenz training data parameters}
\begin{itemize}
	\item dynamical equation: \\
	$\dot{x}_1 = \sigma(x_2-x_1) \hspace{3cm} \dot{x}_2 = x_1(\rho-x_3)-x_2 \hspace{3cm} \dot{x}_3 = x_1x_2 - \beta x_3$.
	\item parameters: $\sigma = 10, \beta = 8/3, \rho = 28$ (except in the bifurcation example).
	\item uniform random initial conditions: $x_1,x_2,x_3 \in [0,10]$.
	\item throwaway simulation time (per attractor): $T_{\mathrm{waste}} = 20$.
	\item training simulation time (per attractor): $T_{\mathrm{train}} = 200$.
	\item translation training shift: $P = \begin{bmatrix} 1\\0\\0\end{bmatrix}$.
	\item transformation training stretch: $P = \begin{bmatrix} -.012 & 0 & 0\\ 0 & 0 & 0\\ 0 & 0 & 0 \end{bmatrix}$.
\end{itemize}

\subsubsection{Reservoir}
\begin{itemize}
	\item dynamical equation: $\frac{1}{\gamma} \dot{\bm{r}} = -\delta\bm{r} + U(A\delta\bm{r}+B\bm{x}+C\bm{c}) + V(A\delta\bm{r}+B\bm{x}+C\bm{c})^2$.
	\item reservoir initial condition: $\bm{r}(0) = \bm{0}$.
	\item adjacency matrix $A$ has 10\% binary density. We begin with matrix $\tilde{A}$ where each nonzero element is drawn from a uniform random distribution $\tilde{A}_{ij} \in [-1, 1]$. Then the matrix is normalized such that $A = 0.95 \frac{\tilde{A}}{\mathrm{real}(\lambda_{\mathrm{max}})}$, where $\lambda_{\mathrm{max}}$ is the eigenvalue with the largest real value. For a $tanh$ activation function, this normalization is used to ensure the echo-state property in discrete time reservoir systems \cite{Jaeger2010}.
	\item number of reservoir neurons: $N = 300$.
	\item time constant: $\gamma = 100$.
	\item fixed point: each element was drawn from a random uniform distribution $r_i^* \in [-1,-0.8] \cup [0.8,1]$.
	\item data input matrix: every row $i$ of $B$ has one non-zero element at index $j$. This index is chosen uniformly at random ($Pr(j=1) = \dotsm = Pr(j=M)$). For examples involving translations and transformations, the value of the element is drawn uniformly from $B_{ij} \in [-0.004,0.004]$. For examples involving bifurcations, the magnitude of the observed data is much smaller (local to the stable fixed point, instead of the full chaotic attractor manifold), and therefore the element is drawn uniformly from $B_{ij} \in [-0.04,0.04]$.
	\item control input matrix: every row $i$ of $C$ has one non-zero element at index $j$. This index is chosen uniformly and randomly ($Pr(j=1) = \dotsm = Pr(j=K)$), and the value of the element is drawn uniformly from $C_{ij} \in [-0.002,0.002]$.
\end{itemize}

\subsection{Simulation method}
To simulate both the input and reservoir dynamics, we used a 4-th order Runge-Kutta numerical integration. For the dynamics of the Lorenz attractor,
\begin{align*}
\dot{\bm{x}} = \bm{f}(\bm{x}),
\end{align*}
the Runge-Kutta computes the following values
\begin{align*}
k_{x1} &= \Delta t \cdot \bm{f}\left(\bm{x}(t)\right)\\
k_{x2} &= \Delta t \cdot \bm{f}\left(\bm{x}(t) + \frac{k_{x1}}{2}\right)\\
k_{x3} &= \Delta t \cdot \bm{f}\left(\bm{x}(t) + \frac{k_{x2}}{2}\right)\\
k_{x4} &= \Delta t \cdot \bm{f}\left(\bm{x}(t) + k_{x3}\right),
\end{align*}
and evolves the state forward using
\begin{align*}
\bm{x}(t+\Delta t) = \bm{x}(t) + \frac{1}{6}(k_{x1} + 2k_{x2} + 2k_{x3} + k_{x4}).
\end{align*}

The simulation of the reservoir dynamics requires more careful analysis, because it is a system driven by external inputs. For the original reservoir dynamics given by Eq.~\ref{eq:quadratic}
\begin{align*}
\frac{1}{\gamma}\dot{\bm{r}} = \bm{f}(\bm{r},\bm{x},\bm{c}) = -\delta \bm{r} + U(A\delta\bm{r} + B\bm{x} + C\bm{c}) + V(A\delta\bm{r} + B\bm{x} + C\bm{c})^2,
\end{align*}
the algorithm to update the reservoir states is given by
\begin{align*}
k_{r1} &= \Delta t \cdot \bm{f}\left(\bm{r}(t),\bm{x}(t),\bm{c}(t)\right)\\
k_{r2} &= \Delta t \cdot \bm{f}\left(\bm{r}(t) + \frac{k_{r1}}{2}, \bm{x}(t) + \frac{k_{x1}}{2}, \bm{c}(t) + \frac{k_{c1}}{2}\right)\\
k_{r3} &= \Delta t \cdot \bm{f}\left(\bm{r}(t) + \frac{k_{r2}}{2}, \bm{x}(t) + \frac{k_{x2}}{2}, \bm{c}(t) + \frac{k_{c2}}{2}\right)\\
k_{r4} &= \Delta t \cdot \bm{f}\left(\bm{r}(t) + k_{r3}, \bm{x}(t) + k_{x3}, \bm{c}(t) + k_{c3}\right),
\end{align*}
and the reservoir state evolves forward according to
\begin{align*}
\bm{r}(t+\Delta t) = \bm{r}(t) + \frac{1}{6}(k_{r1} + 2k_{r2} + 2k_{r3} + k_{r4}).
\end{align*}
Hence, when we simulate the Lorenz state $\bm{x}(t)$, we also save the corresponding values $k_{x1},\dotsm,k_{x3}$ to use in the reservoir update algorithm. Finally, we note that in our simulations, we slowly vary the control input $\bm{c}(t)$ over time, requiring us to determine the trajectory of $\bm{c}(t)$ beforehand. However, we require a differential equation that generated $\bm{c}(t)$ to solve for the final parameters $k_{c1},\dotsm,k_{c4}$. We assume the differential equations that generate $\bm{c}$ are constant, such that between time $t$ and $t+\Delta t$, the rate of change of $\bm{c}(t)$ is given by
\begin{align*}
\dot{\bm{c}}(t) = \bm{f}(\bm{c}(t)) = \frac{\bm{c}(t+\Delta t) - \bm{c}(t)}{\Delta t}.
\end{align*}
Such dynamics yield the parameters
\begin{align*}
k_{c1} &= \Delta t \cdot \bm{f}\left(\bm{c}(t)\right) = \bm{c}(t+\Delta t) - \bm{c}(t)\\
k_{c2} &= \Delta t \cdot \bm{f}\left(\bm{c}(t) + \frac{k_{c1}}{2}\right) = \Delta t \cdot \bm{f}\left(\frac{\bm{c}(t+\Delta t) + \bm{c}(t)}{2}\right) = \bm{c}(t+\Delta t) - \bm{c}(t)\\
k_{c3} &= \Delta t \cdot \bm{f}\left(\bm{c}(t) + \frac{k_{c2}}{2}\right) = \Delta t \cdot \bm{f}\left(\frac{\bm{c}(t+\Delta t) + \bm{c}(t)}{2}\right) = \bm{c}(t+\Delta t) - \bm{c}(t).
\end{align*}
The same integration is used with feedback where $\frac{1}{\gamma}\dot{\bm{r}} = \bm{f}(\bm{r},\bm{c}) = -\delta\bm{r} + U(R\delta{\bm{r}} + C\bm{c}) + V(R\delta{\bm{r}} + C\bm{c})^2$.
\newpage

\subsection{Training}
\label{sec:training}
Using the dynamical equations and RK4 integration scheme, we first generated the Lorenz attractor training inputs $\bm{x}(t)$. Each of the single direction translation and transformation examples described in the main text used four Lorenz attractor inputs. The first was the original Lorenz time series $\bm{x}(t)$, and the remaining three were translations or rotations of the original. Each of these four time series were simulated for $T = T_{\mathrm{waste}} + T_{\mathrm{train}} = 20 + 200 = 220$ time. At a time step of $\Delta t = 0.001$, each time series $\bm{x}(t)$ contained $\frac{T}{\Delta t} = 220,000$ simulation time points, stored in data matrix $X_0$ for the original attractor. Because we also kept the 4 outputs of the $RK4$ numerical integration scheme, the data matrix $X_0$ had dimensions $\mathrm{variables} \times \mathrm{time~steps} \times \mathrm{RK4} = 3 \times 220,000 \times 4$. With three additional time series for translation or rotation, $X_1,X_2,X_3$, we concatenated the four time series along the second dimension into the full matrix $X$ with dimension $3 \times 880,000 \times 4$. 

Using this Lorenz data matrix $X$, and a corresponding control input data matrix, we drove the reservoir to generate $\bm{r}(t)$, contained in a reservoir data matrix $D$ that was of size $N=300 \times 880,000$. For every $\frac{T}{\Delta t} = 220,000$ time steps, we threw away the first $\frac{T_{\mathrm{waste}}}{\Delta t} = 20,000$ time points, as this simulation allowed both the Lorenz and reservoir systems to forget their initial conditions. The remaining $\frac{T_{\mathrm{train}}}{\Delta t} = 200,000$ time points of each attractor were kept for training. This process yields a Lorenz training matrix $X_{\mathrm{train}}$ of dimension $3 \times 800,000$ (as we throw away the RK4 simulation parameters after driving the reservoir), and a reservoir training matrix $D_{\mathrm{train}}$ of dimension $300 \times 800,000$.

Finally, we seek a training matrix $M$ of dimension $3 \times 300$ that minimizes the matrix 2-norm
\begin{align*}
\|M D_{\mathrm{train}} - X_{\mathrm{train}}\|_2.
\end{align*}
Specifically, we use MATLAB's command \verb|lsqminnorm|, that not only minimizes this norm, but in the event that multiple solutions exist, also minimizes the norm of $M$. 
\newpage

\subsection{Training maps the fixed point to 0}
Here, we describe a particular but important methodological nuance to the training. Recall that from Eq.~\ref{eq:quadratic}, our reservoir evolves according to
\begin{align*}
\frac{1}{\gamma}\delta\dot{\bm{r}} = -\delta \bm{r} + U(A\delta\bm{r} + B\bm{x} + C\bm{c}) + V(A\delta\bm{r} + B\bm{x} + C\bm{c})^2,
\end{align*}
where $\delta\bm{r} = \bm{r} - \bm{r}^*$. If our training scheme involved approximating the inputs with the \emph{difference} of the reservoir states $\delta\bm{r}$ such that $W\delta\bm{r}(t) \approx \bm{x}(t)$, then this nuance would be unnecessary, as the feedback dynamics would take the proper form
\begin{align*}
\frac{1}{\gamma}\dot{\bm{r}}' 
&= -\delta \bm{r}' + U(A\delta\bm{r}' + BW\delta\bm{r}' + C\bm{c}) + V(A\delta\bm{r}' + BW\delta\bm{r}' + C\bm{c})^2\\
&= -\delta \bm{r}' + U(R\delta\bm{r}' + C\bm{c}) + V(R\delta\bm{r}' + C\bm{c})^2.
\end{align*}
Unfortunately, such a scheme would require the additional assumptions that during training, the neural system was able to accurately retain knowledge of its fixed point $\bm{r}^*$, and that it was also able to take the difference of the neural activity with respect to this fixed point in real time. We avoid these additional assumptions by training on the true reservoir states such that $W\bm{r}(t) \approx \bm{x}(t)$, yielding
\begin{align*}
\frac{1}{\gamma}\delta\dot{\bm{r}}' 
&= -\delta \bm{r}' + U(A\delta\bm{r}' + BW\bm{r}' + C\bm{c}) + V(A\delta\bm{r}' + BW\bm{r}' + C\bm{c})^2\\
&= -\delta \bm{r}' + U(R\delta\bm{r}' + BW\bm{r}^* + C\bm{c}) + V(R\delta\bm{r}' + BW\bm{r}^* + C\bm{c})^2.
\end{align*}
In both the linear and quadratic terms, we notice an extra and undesired term $BW\bm{r}^*$. In all of our simulations, the training of matrix $W$ actually maps the fixed point to a small number, such that $W\bm{r}^*$ is on the order of $10^{-6}$, whereas $W\delta\bm{r}(t)$ is on the order of $10^1$. Hence, the matrix $W$ maps the fixed point to values that are 7 orders of magnitude smaller than the magnitude of the inputs, such that $W\bm{r}'(t) = W\delta\bm{r}'(t) + W\bm{r}^* \approx W\delta\bm{r}'(t)$, thereby rendering the undesired term $BW\bm{r}^*$ effectively negligible.

At first, we might be tempted to explain this phenomenon by the fact that the Lorenz attractor $\bm{x}(t)$ is centered around $x_1 = 0$ and $x_2 = 0$. Hence, it would make sense that a constant fixed point $\bm{r}^*$ would map to a value of $0$. However, the third coordinate of the Lorenz system is centered around $x_3 = \rho-1 = 27$, and yet in our simulations training still produces an output matrix M that maps the fixed point to 0, even along the $x_3$ coordinate, across randomly assigned fixed points $\bm{r}*$ and reservoir parameters $A$, $B$, and $C$.
\newpage

\subsection{Truncation of the block-Hessenberg matrix}
To understand the mechanism of learning translations and transformations, we had taken the differential of the reservoir feedback dynamics,
\begin{align*}
(I-KA)d\bm{r}' + \frac{1}{\gamma}d\dot{\bm{r}}' = K(BWd\bm{r}' + Cd\bm{c}).
\end{align*}
If we take time derivatives of the left-hand side of this equation, we obtain
\begin{align*}
\begin{bmatrix}
(I-KA) & & \frac{1}{\gamma}I & & 0 & & 0 & & 0 & \dotsm\\
\downarrow & \searrow & & \searrow\\
-\dot{K}A & & (I-KA) & & \frac{1}{\gamma}I & & 0 & & 0 & \dotsm\\
\downarrow & \searrow & \downarrow & \searrow & & \searrow\\
-\ddot{K}A & & -2\dot{K}A & & (I-KA) & & \frac{1}{\gamma}I & & 0 & \dotsm\\
\downarrow & \searrow & \downarrow & \searrow & \downarrow & \searrow & & \searrow\\
-\dddot{K}A & & -3\ddot{K}A & & -3\dot{K}A & & (I-KA) & & \frac{1}{\gamma}I & \dotsm\\
\vdots & & \vdots & & \vdots & & \vdots & & \vdots & \ddots
\end{bmatrix}
\begin{bmatrix}
d\bm{r}'\\ \\
d\dot{\bm{r}}'\\ \\
d\ddot{\bm{r}}'\\ \\
d\dddot{\bm{r}}'\\
\vdots
\end{bmatrix},
\end{align*}
where the element in the $i$-th row and $j$-th column has a coefficient 
\begin{align*}
p_{i,j} = \begin{pmatrix}i-1\\j-1\end{pmatrix} \hspace{1cm} \mathrm{for}~j \leq i,
\end{align*}
according to Pascal's triangle. For the translation examples, we can write the continued time derivatives of the differential relation as
\begin{align*}
\underbrace{
\begin{bmatrix}
H_0 & H_{-1} & 0 & 0 & \dotsm\\
H_1 & H_0 & H_{-1} & 0 & \dotsm\\
H_2 & 2H_1 & H_0 & H_{-1} & \dotsm\\
H_3 & 3H_2 & 3H_1 & H_0 & \dotsm\\
\vdots & \vdots& \vdots & \vdots & \ddots
\end{bmatrix}}_{J}
\begin{bmatrix}
d\bm{r}'\\
d\dot{\bm{r}}'\\
d\ddot{\bm{r}}'\\
d\dddot{\bm{r}}'\\
\vdots
\end{bmatrix}
\approx
\begin{bmatrix}
K\\
\dot{K}\\
\ddot{K}\\
\dddot{K}\\
\vdots
\end{bmatrix}
(BP + C)d\bm{c},
\end{align*}
where $H_{-1} = \frac{1}{\gamma}I$, $H_0 = I-KA$, and $H_i = -K^{(i)}A$ is the $i$-th time-derivative of $KA$. This matrix is a block matrix (each element $H$ is a matrix), and is specifically a block-Hessenberg matrix (zero above the first block-super diagonal). The goal here is to solve for $d\bm{r}'$ with respect to $d\bm{c}$. If we truncate $J$ to a finite-dimensional matrix such that
\begin{align*}
J \simeq 
\begin{bmatrix}
J_{11} & J_{12}\\
J_{21} & J_{22}
\end{bmatrix},
\end{align*}
where
\begin{align*}
J_{11} &= 
\begin{bmatrix}
H_0\\H_1\\\vdots\\H_{k-1}
\end{bmatrix},
\hspace{2cm}
J_{12} = 
\begin{bmatrix}
H_{-1}, & 0, & \dotsm, & 0\\
H_0, & H_{-1}, & \dotsm, & 0\\
\vdots & \vdots & \ddots & \vdots\\
p_{k,2}H_{k-2}, & p_{k,3}H_{k-3}, & \dotsm, & H_{-1}
\end{bmatrix},\\
J_{21} &= 
\begin{bmatrix}
p_{k+1,1}H_k
\end{bmatrix},
\hspace{1.6cm}
J_{22} = 
\begin{bmatrix}
p_{k+1,2}H_{k-1} & p_{k+1,3}H_{k-2} & \dotsm & H_0
\end{bmatrix},
\end{align*}
Then, the closed form solution for the first $N$ rows of $J^{-1}$ (the first block) can be written \cite{Sowik2018} as
\begin{align}
\label{eq:HTBM}
[J^{-1}]_{(1:N,:)} \simeq 
-(J_{22} J_{12}^{-1} J_{11} - J_{21})^{-1}
\begin{bmatrix}
-J_{22} J_{12}^{-1} & I
\end{bmatrix}.
\end{align}
However, in reality, $J$ is not a finite matrix, but an infinite dimensional matrix. An important fact to verify, then, is whether there exists a sufficiently large value of $k$ to yield an accurate inversion. While proving that this inverse converges is outside the scope of this work, we numerically demonstrate in what follows that after $k=1$, successive terms do not perceivably change the results. Specifically, we solve for $d\bm{r}'$ with respect to $d\bm{c}$ for $k = 0, 1, 2, 3$.

\begin{figure}[h!]
	\includegraphics[width=6.5in]{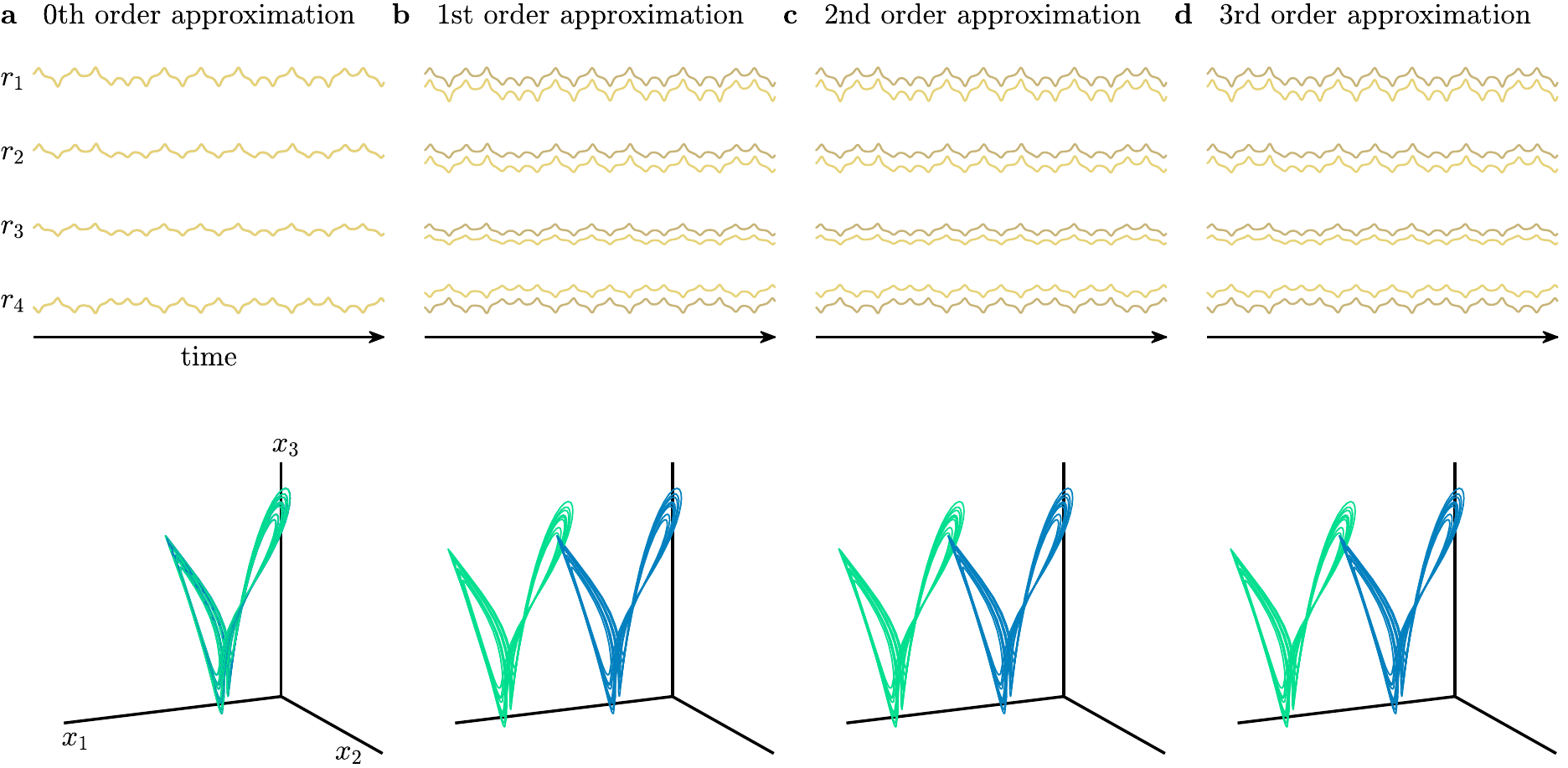}
	\caption{\textbf{Predicted change in reservoir states given a change in control parameter.} Reservoir time series generated by driving the reservoir with the original Lorenz input with $c = 0$ (dark gold), and the predicted time series from solving for $d\bm{r}'$ after training on translated examples and changing the control parameter $\Delta c = 20$ (light gold), along with their output projections (dark and light red, respectively). These approximations were taken by computing the inverse Eq.~\ref{eq:HTBM} for (\textbf{a}) $k=0$, (\textbf{b}) $k=1$, (\textbf{c}) $k=2$, and (\textbf{d}) $k=3$.}
	\label{fig:sf3}
\end{figure}

As a reference for translation, at $k=0$, the approximation becomes
\begin{align*}
d\bm{r}' \approx H_0^{-1} K(BP+C)d\bm{c},
\end{align*}
and at $k=1$, we obtain the approximation used in the main text. The $0$-th order approximation at $k=0$ yields no change (Fig.~\ref{fig:sf3}a), where the predicted reservoir states (light gold) are identical to the original states (dark gold). The first order approximation at $k=1$ (Fig.~\ref{fig:sf3}b) yields a change in the reservoir states that outputs to the expected translation in spatial coordinates. Taking more terms in the approximation ($k=2$, Fig~\ref{fig:sf3}c; and $k=3$, Fig.~\ref{fig:sf3}d) yields no perceivable change in either the reservoir states or their outputs.

\newpage

\subsection{Summary}
In sum, we have provided a general form for reservoir dynamics (Eq.~\ref{eq:idynamics}), the derivation for the quadratic form of the reservoir (Eq.~\ref{eq:quadratic}), as well as the dependence of matrices $U$ and $V$ that arise from the choice of fixed point when using $tanh$ as the activation function (Eq.~\ref{eq:U},\ref{eq:V}). We further provide all simulation parameters (time step, Lorenz parameters and initial conditions, reservoir parameters and initial conditions), along with the specific details of our simulation method, data dimensions, and training process. Finally, we provide numerical justification for the truncation of our approximation when deriving the mechanism of learning (Eq.~\ref{eq:HTBM}).
\newpage

\section{Results}
In this section, we provide some additional results to support the generalizability of the framework.

\subsection{Translation in multiple directions}
\begin{figure}[h!]
	\includegraphics[width=6.5in]{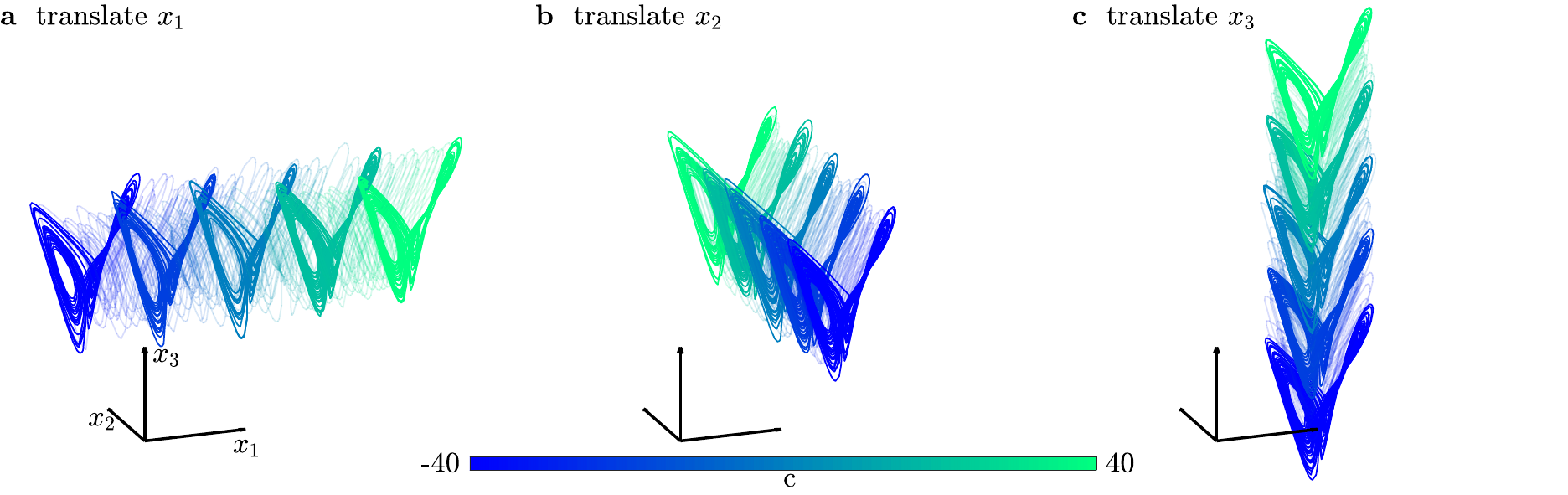}
	\caption{\textbf{Translation of the Lorenz representation in all three spatial directions.} (\textbf{a}) Output of the feedback reservoir after being trained on 4 time series of a Lorenz attractor translated in the $x_1$ direction at $c = 0,\dotsm,4$. By varying $c$ from $-40$ to $40$, the representation shifts in the $x_1$ direction. (\textbf{b}) The same scheme is employed for translations in the $x_2$ direction, and (\textbf{c}) in the $x_3$ direction.}
	\label{fig:sf1}
\end{figure}

In the main text, we demonstrated that a reservoir can translate its representation of a Lorenz attractor along the $x_1$ direction. Specifically, we took an untranslated Lorenz time series $\bm{x}_0(t)$, and generated three additional training examples $\bm{x}_c(t)$ for $c = 1,2,3$ such that
\begin{align*}
\bm{x}_c(t) = \bm{x}_0(t) + 
c\begin{bmatrix}
1\\0\\0
\end{bmatrix}.
\end{align*}
We then drove the reservoir using these four training examples and an additional control parameter $c$ that we also varied from $c = 0, \dotsm, 4$. Afterwards, we performed the feedback, and translated the reservoir's representation by varying the external control parameter $c$ from $-40$ to $40$. We reproduce this translated representation here (Fig.~\ref{fig:sf1}a). We show the same output of the feedback reservoir trained on four examples translated in the $x_2$ direction ($\bm{x}_c(t) = \bm{x}_0(t) + c[0;1;0]$) and in the $x_3$ direction ($\bm{x}_c(t) = \bm{x}_0(t) + c[0;0;1]$) (Fig.~\ref{fig:sf1}b,c). Hence, we demonstrate that the reservoir can learn these translations in arbitrary directions.
\newpage

\subsection{Different types of transformations}
\begin{figure}[h!]
	\includegraphics[width=6.5in]{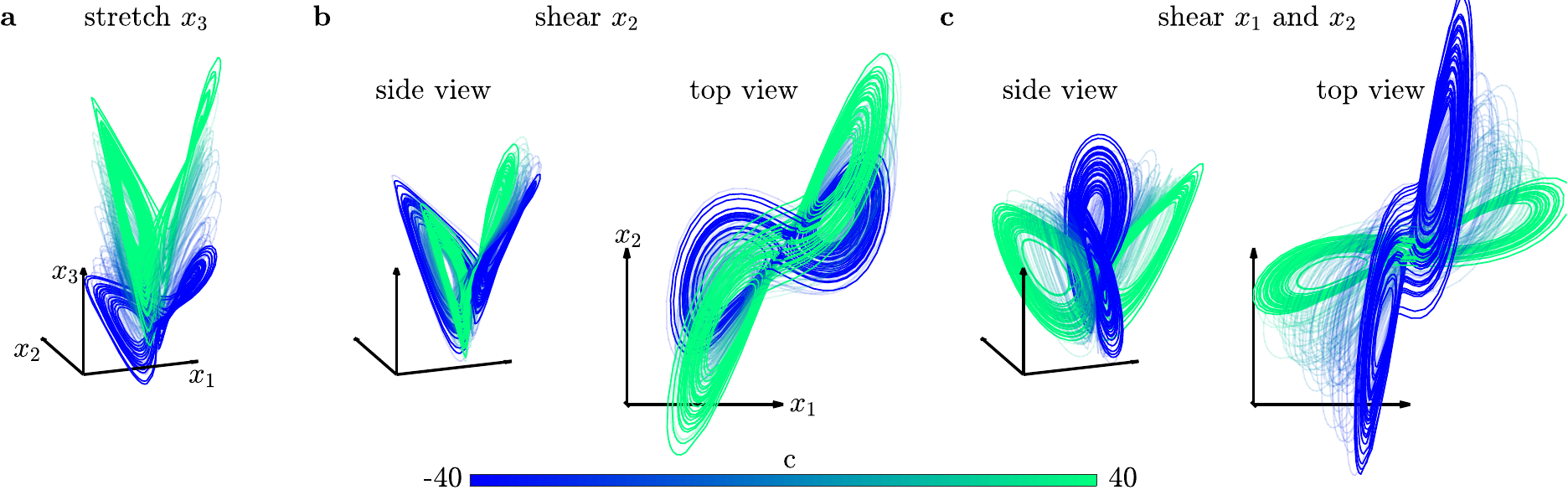}
	\caption{\textbf{Transformation of the Lorenz representation using stretch and shear in several spatial directions.} (\textbf{a}) Output of the feedback reservoir after being trained on 4 time series of a Lorenz attractor stretched in the $x_3$ direction at $c = 0,\dotsm,4$. By varying $c$ from $-40$ to $40$, the representation stretches in the $x_3$ direction. (\textbf{b}) The same scheme is employed for a shear in the $x_2$ direction, and (\textbf{c}) for a shear in the $x_1$ and $x_2$ directions.}
	\label{fig:sf2}
\end{figure}

In the main text, we demonstrated that a reservoir trained on the original Lorenz attractor $\bm{x}_0(t)$ and on three transformed examples $\bm{x}_c(t) = (I+cP)\bm{x}_0(t)$ for $c = 1,2,3$, was able to continuously interpolate and extrapolate the transformation on its internal representation, even for control inputs between $-40$ and $40$. Here, we consider a stretch in the $x_3$ direction, a shear in the $x_1$ direction, and a shear in the $x_1$ and $x_2$ directions. Specifically, we use the three matrices
\begin{align*}
P_{\mathrm{stretch,x_3}} = 
\begin{bmatrix}
0 & 0 & 0\\
0 & 0 & 0\\
0 & 0 & 0.012
\end{bmatrix},
\hspace{.7cm}
P_{\mathrm{shear,x_1}} = 
\begin{bmatrix}
0 & 0 & 0\\
0.012 & 0 & 0\\
0 & 0 & 0
\end{bmatrix},
\hspace{.7cm}
P_{\mathrm{shear,x_1,x_2}} = 
\begin{bmatrix}
0 & -0.012 & 0\\
0.012 & 0 & 0\\
0 & 0 & 0
\end{bmatrix},
\end{align*}
to train our reservoir for $c = 0,\dotsm,4$. For each transformation, we drive the reservoir with the input Lorenz attractors $\bm{x}_c(t)$ and an additional control input $c$ for $c = 0,\dotsm,4$. We then train the reservoir by applying the feedback method used in the main text. Finally, we drive the autonomous feedback reservoir by varying the control parameter from $c = -40$ to $c = 40$ for these three transformations (Fig.~\ref{fig:sf2}).
\newpage

\newpage
\bibliography{references}